# Global coupling at 660 km is proposed to explain plate tectonics and the generation of the earth's magnetic field

Jozsef Garai

*Department of Earth Sciences, Florida International University, Miami, FL 33199, USA*

*E mail: jozsef.garai@fiu.edu*

The presence of low viscosity layers in the mantle is supported by line of geological and geophysical observations. Recent high pressure and temperature investigations indicated that partial carbonate melt should exist at the bottom of the lithosphere and at 660 km. The presence of few percent carbonate melt reduces the viscosity by several order of magnitude. The globally existing 660 km very low viscosity layer allows the development of differential rotation between the upper and lower mantle. This differential rotation between the 660 km outer shell and the rest of the earth offers a plausible explanation for plate tectonics and for the generation of the earth's magnetic field. Simple dynamo model is proposed, which able to reproduce all of the features of the contemporary and, within reasonable uncertainty, the paleomagnetic field. The model is also consistent with geological and geophysical observations.

## 1. Introduction

Plate tectonics theory (Vine & Mathews 1963; Vine, 1966), which revolutionized our understanding of the earth and successfully describes the motions of the continents, is incomplete. Despite the extensive research there is no adequate explanation for the causes of the plate motions.

The widely held consensus is that heat related mantle convections provide the energy for the plate movements. This hypothesis can easily be tested. The physics of the convections is known, the motions of the plates are well defined, and the viscosity of the mantle is known with the required accuracy. Using these data allows one to determine global convections pattern. By putting the convection cells under the plates result in many trouble spots signaling problems with the model. The discrepancies could be resolved by reducing the viscosity of the mantle. The lower viscosity would allow more flexibility between the convection cells and plates and the kinematics of the plates could be reproduced. However, the reduction of mantle viscosity results in smaller convection cells with insufficient size. The paradox, high viscosity unable to explain the plate motions and low viscosity unable to produce the required size of the convection cells is mutually exclusive. It is suggested that mantle convection models do not provide a credible explanation for plate tectonics and that it is time to change the paradigm.

The presence of thermal convections is not questioned. However it is suggested that these heat convections are passive and generated by the movements of the plates.

Besides heat the only other available energy that the earth possesses its rotational kinetic energy. Discrediting heat convections is proof by exclusion that plate tectonics is driven by the rotational kinetic energy of the earth. Tidal or earth rotation related models are among the earliest explanations for plate tectonics (Wegener, 1915; 1924).

Global westward drift of the plates has been detected from fixed Antarctic plane reference (Knopoff and Leeds, 1972) and from hot spot reference frame (Richard et al. 1991; Gordon 1995). Tidal drag as a mechanism for displacing the lithosphere has been proposed to explain the observed westward drift (Bostrom, 1971; Knopoff and Leeds, 1972; Moore, 1973). Jordan (1974) discredited this proposal be demonstrating that the viscosity which would allow decoupling between the mantle and the lithosphere should be lower than $10^{11}$ Pas. The present day viscosity estimates for the asthenosphere are much higher ($10^{17}$-$20^{20}$ Pas). Recently, based on nonlinear rheology of the mantle, the mechanical fatigue, and the irreversible down welling of heavier rocks in the mantle, the revitalization of the astronomical origin of the westward drift has been proposed by Scoppola et al. (2006).

## 2. Low viscosity layers in the mantle

Investigating the uplift data of the Angerman river it has been suggested that the very rapid uplift occurred right after the deglaciation can be explained only if the viscosity of the asthenosphere is between $10^{18}$ Pas and $10^{11}$ Pas (Garai 1997, 2003). Based on this low viscosity it was suggested that plate tectonics could be driven by the angular momentum of the earth (Garai 1997).

Laboratory high pressure and temperature experiments indicates that San Carlos olivine, the major constituent of the mantle, is not stable in the presence of carbon and decomposes to produce olivine with lower Fe content, pyroxene, carbide, and carbonate melt (Garai and Gasparik, 2000; 2003). Based on the correlation of C and $^{36}$Ar in chondritic meteorites, there is a sufficient amount of carbon in the mantle (Wood et al. 1996; Molina, and Poli 2000) to produce a 0.5-3 percent carbonate melt, observed in the experiments. The presence of carbonate melt modifies the subsolidus phase relations for $(Mg,Fe)_2SiO_4$ in a way that provides a much better agreement with the seismic observations (Fig. 1).

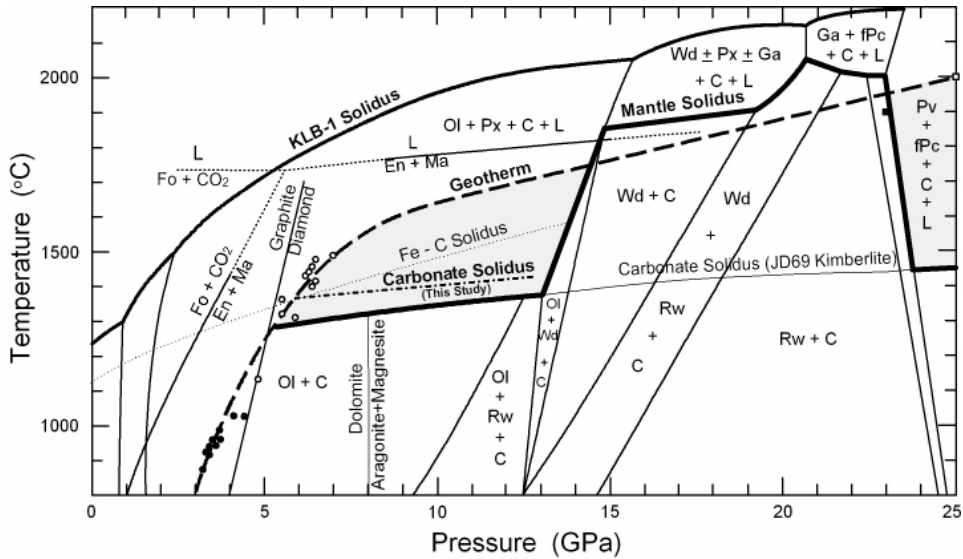

**Fig. 1** Temperature-pressure phase diagram for the composition $(Mg_{0.9}Fe_{0.1})_2SiO_4$. The geotherm was constrained at low pressures by the thermobarometry (Gasparik, 2000) applied to mantle xenoliths from Northern Lesotho (Nixon and Boyd, 1973), and at the lower-mantle pressures by the thermobarometry (Gasparik and Hutchison, 2000) applied to the NaPx-En (Wang and Sueno, 1996) and the Type III (Hutchison, 1997) inclusions in diamonds. Symbols: C=diamond; Ga = garnet; L = liquid/melt; Ol = olivine; $Pv = MgSiO_3$ perovskite; Px = pyroxene; Rw = ringwoodite; Wd = wadsleyite; fPc = ferropericlase, (Mg,Fe)O (Figure from Garai and Gasparik, 2003)

The experimentally determined solidus of the anhydrous peridotite KLB-1 (Herzberg, 2000; Zhang and Herzberg, 1994) relevant for such a mantle, is hundreds of degrees higher than the expected temperatures of an average mantle. In contrast, the presence of partial melt in the low velocity zone (Gutenberg 1948) and possibly in the depth range of 500-1000 km (Kido 1997; Cadek 1998; Montagner 1998) is suggested by the detected high attenuation of seismic waves.

Carbonate melt could reduce the viscosity of the olivine rich solid mantle by more than ten orders of magnitude. Conservative estimations of the experimental results indicate that the presence of about 2-3 percent carbonate melt reduced the apparent viscosity of the olivine-carbonate melt system to $10^8$ Pas. Carbonate could separate through geological times and become trapped in higher concentration at the bottom of the lithosphere and at the bottom of the 660 km boundary.

The low viscosity layer at 660 km is a global phenomenon because it is not disrupted by subducting slabs. It is suggested that this globally existing layer should allow differential rotation between the upper 660 km and the rest of the earth. It is also suggested that this differential rotation with slab buoyancy is responsible for plate tectonics (Fig. 2). The existence of differential rotation at 660 km indicates that mantle plumes can not originate deeper than 660 km.

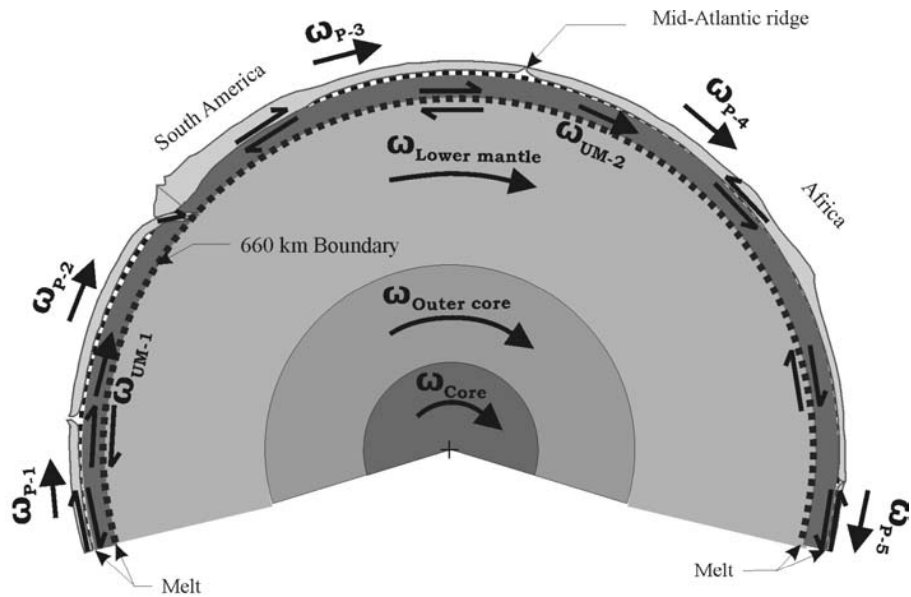

**Figure 2.** Proposed model for plate tectonics. The current rotational gaps are shown. Global coupling or rotational difference is developed at 660 km.

The possible existence of a global differential rotation at 660 km and the consequences are discussed in details.

## 3. Geological evidences supporting low viscosity

Besides postglacial rebounds the presence of low viscosity layer below the lithosphere and at the 660 km boundary is supported by additional observations.

The spreading lines and transform faults are oriented into N-S and E-W direction respectively. 38 % of the total spreading lines have N-S orientation, while 30 % of the total length of the active transform faults is in 10 degree of being parallel with the equator (Moore, 1973). These features are consistent with rotational driven plate movements.

The rotational drag model is energetically feasible because the energy released by tectonic activity ($1.3 \times 10^{19}$ J/yr) is smaller than the energy dissipated by tidal friction ($1.6 \times 10^{19}$ J/ye) (Denis et al. 2002).

Based on the currently accepted viscosity models, the mantle flow in the asthenosphere should maintain a much larger dynamic topography, which is not observed. This is a major unexplained problem in geophysics (Wheeler, 2000). If the viscosity directly below the lithosphere is decreased by several orders of magnitude the amplitude of the dynamic topography on the surface of the Earth is reduced, but only by ~20% (Lithgow-Bertolonni, 1997). A more dramatic decrease in the viscosity of the low viscosity layer might explain the complete lack of the dynamic topography.

The global distribution of the plate velocities and seismic activities tends to decrease toward the poles (DeMets et al. 1990; Heflin, 2004).

Venus with more intense volcanic activity than earth does not show features of plate tectonics. On the other hand the angular momentum of the planet is very small.

Back arch spreading (e.g.. Uyeda and H. Kanamori, 1979) can be easily explained by the coupling at 660 km (Fig. 3).

a./ $\omega_{P1} > \omega_{P2}$

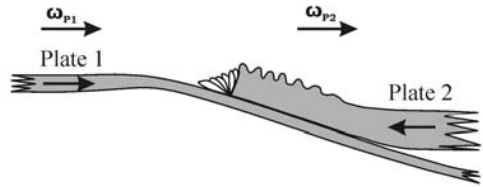

b./ $\omega_{P1} < \omega_{P2}$

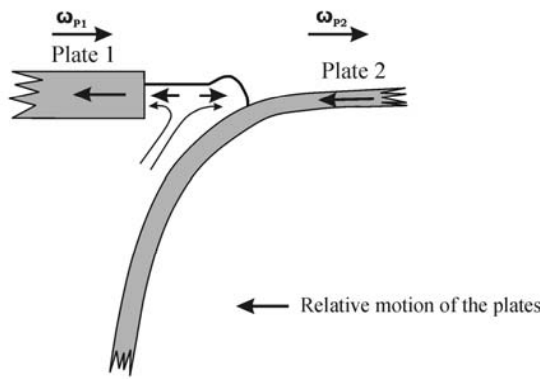

(Latitudinal cross sections viewed from South to North)

**Figure 3.** The angular velocity difference of the plates able to explain all of the features of the subduction zones. (a) the preceding plate rotate slower (b) the preceding plate rotates faster (back arch spreading)

It has been suggested that the sign of the angular velocity difference at 660 km can be changed (Garai 2001). Based on the penetration profile of the subduction slabs it is suggested that currently the angular velocity of the lower mantle is higher than angular velocity of the upper mantle. Slabs subducting from east to west are flattened out at 660 km and trapped above the 660 km discontinuity (Fukao et al. 1992). This flattening can be explained by the faster rotation of the lower mantle. On the other hand slabs subducting from west to east ends up sharply at 660 km with no flattening. Slabs subducting from north to south or south to north are positioned parallel to the mantle flow with no mantle pressure on either sides. The minimum of 1500 km penetration of the Sunda slab below Java (Widiyantoro and van der Hilst, 1996) can be explained by its parallel position to the mantle flow.

The proposed east-west coupling is also supported by the observed melt production under the East Pacific Rise which is asymmetrical towards to the west (MELT 1998; Toomey et al. 1998).

## 4. Implication to the earth's magnetic field

There is a consensus among geophysicists that the earth's magnetic field is generated by the fluid motion of the liquid iron core, which maintained by heat and gravitational energy from the cooling earth (e.g. Merill and McElhinny, 1983; Roberts and Glatzmaier, 2000). Besides the considerable effort has been made there is no magnetohydrodynamics model which is able to explain even the basic observed features of the field, like the reversals. The proposed differential rotation at 660 km offers plausible explanation for the earth's magnetic field (Garai, 2001).

The temperature differences inside the earth diffuses electrons (or holes) (Ullmann, 2007; MacDonald and Chalmers, 1962) resulting in a thermally induced inhomogeneous charge distribution. The charge builds up on the cold end and creates electric field inside the earth. The proposed global differential rotation at 660 km should generate magnetic field. Reversing the field would require to change the sign of the differential rotation. The tidal drag affect is continuous and works only one direction; however, mass redistribution on the surface of the earth can also change the angular velocity. If the angular velocity caused by mass redistributions overrides the affect of the rotational drag then the rotational sign at 660 km can be changed resulting in the reversal of the field.

## 5. Characteristics of the field

The earth's magnetic field has been monitored for centuries and the considerable amount of data allows us to characterize the basic features of the field (e.g. McDonald and Gust, 1965). The well established characteristics of the field are:

i.  in geologic timescale the poles of the magnetic fields coincide with the rotational axis of the earth,
ii. the field reverses its polarity,
iii. the current field has normal polarity,
iv. the majority of the field is dipole (current value is around 90%),
v.  the moment of the geocentric axial dipole has been decreasing in the last hundred years at about 0.05% average/year,
vi. both the dipole and the non-dipole field drifts westerly,
vii. the drift velocity of the dipole field during the last 150 years is about 0.05 degree/year,
viii. the nondipole field moved faster during the same period with the drift velocity of about 0.18 degree/yr (e.g. Bullard et.al., 1950),
ix. periodicities related to solar and lunar (Kreil, 1850) cycles are identified in the strength of the magnetic field,
x.  the virtual geomagnetic poles (VGP) during reversals favor two particular rather narrow antipodal bands. One passes through Americas while the other through Central Asia and western Australia (e.g., Tric et al., 1991; Clement, 1991; Laj et al. 1991),
xi. the VGP seems to prefer the "American path" during N-R reversals, while during R-N reversals moves through the "Asian path" (Gubbins and Coe, 1993),
xii. the polarity transition sometimes very fast. A few tenth of degrees change of the direction of the field was detected in less than one year. The intensity of the field also underwent rapid changes (Prevot et al., 1985; Coe and Prevor, 1989) check
xiii. the rate of the westward drift varies with time and seems to associate with the fluctuations in the length-of-day (LOD) (Vestine, 1952),
xiv. the decade variations of the geomagnetic field and the LOD show similar patterns in their variation. (e.g., Vestine and Kahle, 1968) The sectorial components of the geomagnetic decadal variations also correlate well with LOD variation. (e.g., Yoshida and Hamano, 1995),

xv. the intensity and the inclination of the Earth's magnetic field reveal the presence of 100,000-year periodicity indicating that the field is modulated by the orbital eccentricity (Yamazaki and Oda, 2002),

xvi. the correlation between inclination and intensity shifts from antiphase to in-phase when the polarity changes from reversed to normal (Yamazaki and Oda, 2002),

xvii. the changes of the geomagnetic field correlates well with the gravitational fields when one field is displaced in longitude by a certain angle (Hide and Malin, 1970, 1971,a,b),

xviii. the last two R-N magnetic reversals, the Jaramillo and the Brunhes-Matuyama are associated with microtektite strewnfield indicating major impact at the time of the reversal,

xix. the Brunhes-Matuyama reversal also associated with changes in temperature and biota extinction (Glass, 1982, Glass et al., 1979),

xx. there is an inverse correlation between the frequency of magnetic reversals and plum activity for the past 150 Ma, as measured by the volume production rate of oceanic plateaus, seamount chains, and continental flood basalts (Larson and Olson, 1991),

xxi. the geomagnetic field was stronger during the Cretaceous N and Permian R polarity superchrons than during the superchrons of frequent reversals. The magnetic field of the N polarity superchron had higher intensity in comparison to the R polarity superchron. These superchrons are separated in time by other about 120 Myr (Pal and Roberts, 1988),

xxii. there is a causal relationship between the Earth's magnetic field and climate variation and faunal extinction,

xxiii. Despite the similar size and composition and the presence of molten iron core Venus does not have intrinsic magnetic field or $10^{-5}$ times smaller than the magnetic field of Earth (Luhmann and Russel, 1997).

Any model proposed to explain the generation of the field should explain all of the features of both the current and paleomagnetic field and the correlations of the field to geological and geophysical observations.

Correlation between the earth's magnetic field and the rotation are supported by the line of observations such as i, iv, ix, xiii, xiv, xv, xvi, and xvii. These correlations are indicating that the magnetic field is generated by a rotation related process.

## 6. Proposed model

The simplest rotational model is a dynamo. Assuming a spherical sphere with a radius of R, carrying a uniform surface charge $\sigma$, and spinning at angular velocity $\omega$ gives the magnetic vector potential

$$A(r,\theta,\phi) = \frac{\mu_0 R^4 \omega \sigma}{3} \frac{\sin\theta}{r^2} \hat{\phi} \qquad (r \geq R) \tag{1}$$

where r is the distance from the center of the sphere to a point, $\mu_0$ magnetic permeability of free space, $\theta$ angle of the point from the z axis, and $\phi$ angle of the point with the x axis. The magnetic field [B] can be calculated as:

$$B = \nabla \times A \tag{2}$$

therefore

$$B = \frac{\mu_0 R^4 \omega \sigma}{3} (\frac{2\cos\theta}{r^3} \hat{r} + \frac{\sin\theta}{r^3} \hat{\theta}) \tag{3}$$

The strength of the field [$B_0$] at the equator [$\theta=0$] is

$$B_0 = \frac{\mu_0 R^4 \omega \sigma}{3r^3} \tag{4}$$

It is assumed that the angular velocity of the charged sphere is the equal with the developed angular velocity difference at 660 km [$\omega_i$]. Using the angular velocity of the outer shell (OS) and the lower mantle (LM) gives

$$\omega_i = \Delta\omega = \omega_{LM} - \omega_{OS} \tag{5}$$

It is assumed that the best estimation for this angular velocity difference is equivalent with the westward movement of the dipole field [$0.05^0/y$]. It is also assumed that the charge center of the outer 660 km shell is 200 km below the surface. Using the reference radius for the spherical earth [$6.3712 \times 10^6 m$] the radius of this representative sphere is $R_{OS} = 6.1712 \times 10^6 m$.

The average value of the strength of the current field at the equator can be calculated from the reference model.

$$B_0 = \frac{\mu_0 m}{4\pi r^3} \tag{6}$$

where

$r = 6.3712 \times 10^6 m$

$\mu_0 = 4\pi 10^{-7} kgmA^{-2}s^{-2}$

$m = 7.94 \times 10^{22} \, Am^2$ [dipole moment of the earth's magnetic field]

To generate this strength at the equator with the sphere model $4984 Cm^{-2}$ surface charge is needed. Distributing equally this concentrated surface charge in the entire 660 km, as a first order approximation, the extra charge need to generate the magnetic field of the earth is $6.715 \times 10^{-7} Ccm^3$. The friction generated charge of a glass or a rubber rod is in the same order, therefore, this existing excess charge in the outer 660 km shell of the earth seems to be reasonable. No charge affect for the rest of the earth was taken into account; therefore, the necessary charge to maintain the field is even smaller.

## 7. Testing the model against observations

Changes in the angular velocity of the rotating earth modify the differential rotation between the 660 km outer shell and the rest of the earth and change the strength of the field. This prediction is consistent with the detected correlation between the length of the day (LOD) and the strength of the magnetic field. It is also consistent with the inverse correlation between the westward drift of the field and LOD. Using the spherical model, with the determined surface charge, the long term decrease in the strength of the dipole field, observed between 1600 and 1990, can be generated by an 1.78 ms/cy increase of LOD. This calculated value is consistent with observations. Reproducing the observe decrease of the field from the changes of LOD is an independent proof supporting the correctness of the proposed model.

The detected lunar cycles (Kreil, 1850) in the geomagnetic field can be explained by the secular changes in the rotation of the earth caused by the moon. Contrarily there is no explanation how the moon might have any effect on a core generated magnetic field.

Climate changes by modifying the angular velocity of the earth's rotation through mass redistribution on the surface should have effect the strength of the magnetic field. If the polarity of the field is normal than a glaciation period should increase the strength of the field, while a deglaciation period should vanish the strength of the field. The casual correlation detected between ice ages and the strength of the magnetic field is consistent with the proposed mechanism (Fig. 4).

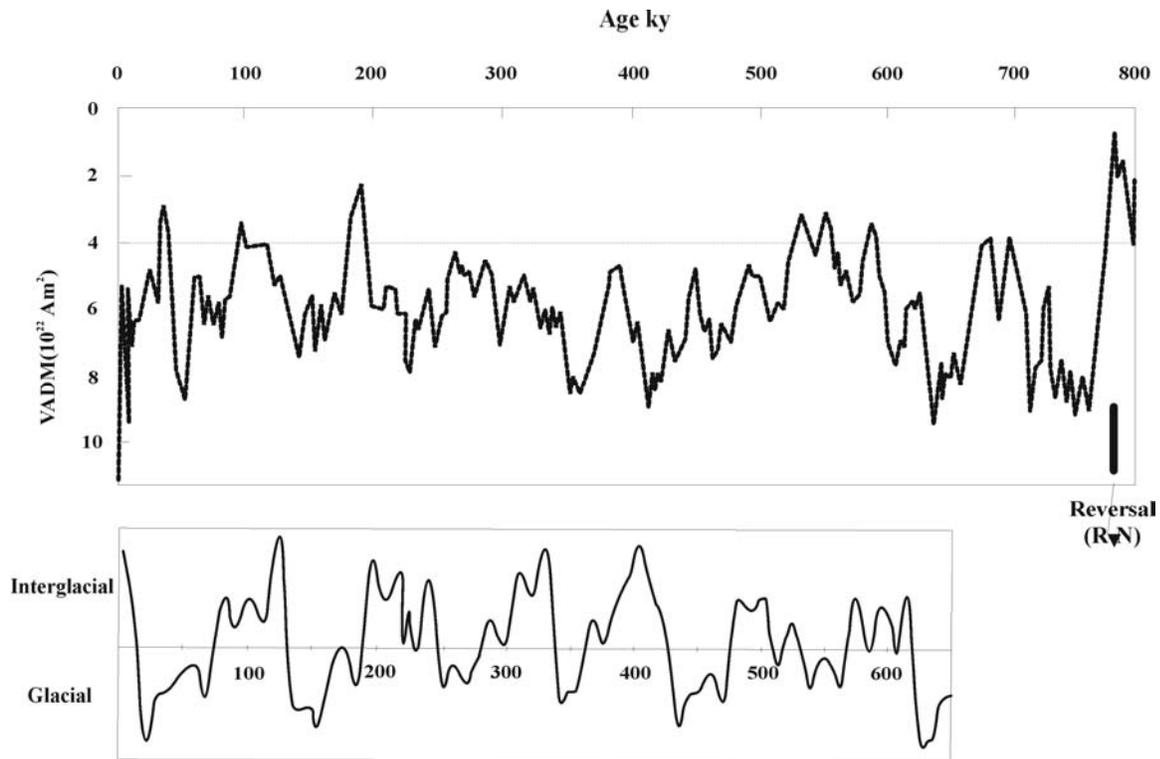

**Figure 4.** Casual correlation between climate and the strength of the magnetic field. The sign of the correlation is consistent with the model. Interglacial period weakens the field while glacial period makes it stronger. The $\delta^{18}O$ curve is from Karner et al. (1996) and the relative variation of the dipole field intensity is from Valet and Meynadier (1993)

Based on the coincidence of climate and magnetic reversals an impact triggered climate change mechanism had been proposed to revert the orientation of the core-generated field (Muller and Morris, 1986). The last two R-N magnetic reversals, the Jaramillo and the Brunhes-Matuyama are associated with microtektite strewnfield which is consistent with an impact initiated field reversal. The Brunhes-Matuyama reversal was also associated with changes in temperature and biota extinction (Glass, 1979, 1982). It is calculated that the Brunhes-Matuyama reversal would require an about 140 m reduction of the global sea level in order to reverse the sign of the differential rotation and generate the observed normal field. The size of this required sea level drop is in the range of the known fluctuation of the eustatic sea level. The Brunhes-Matuyama reversal occurred about 12 kyr after the impact (Schneider et al, 1992), which would provide enough time for the development of a 100 m size of sea level drop. Impact initiated global cooling coinciding with a cold period of the climate cycles seems to be a credible mechanics for N-R reversals.

The sign of the differential rotation between the outer shell and the lower mantle will shift first where the rigidities and the resistance against the flow are the highest. The rigidity of the outer shell is highest at the regions of subduction zones. The subducting slabs have their highest resistance where they subduct against the direction of the differential flow. An East-West subduction would be most effective against a faster rotating LM, while a West-East subduction would be most effective against a slower rotating LM. The observed preferred longitudes of the virtual geomagnetic poles for N-R and R-N reversals are consistent with this prediction (Fig. 5).

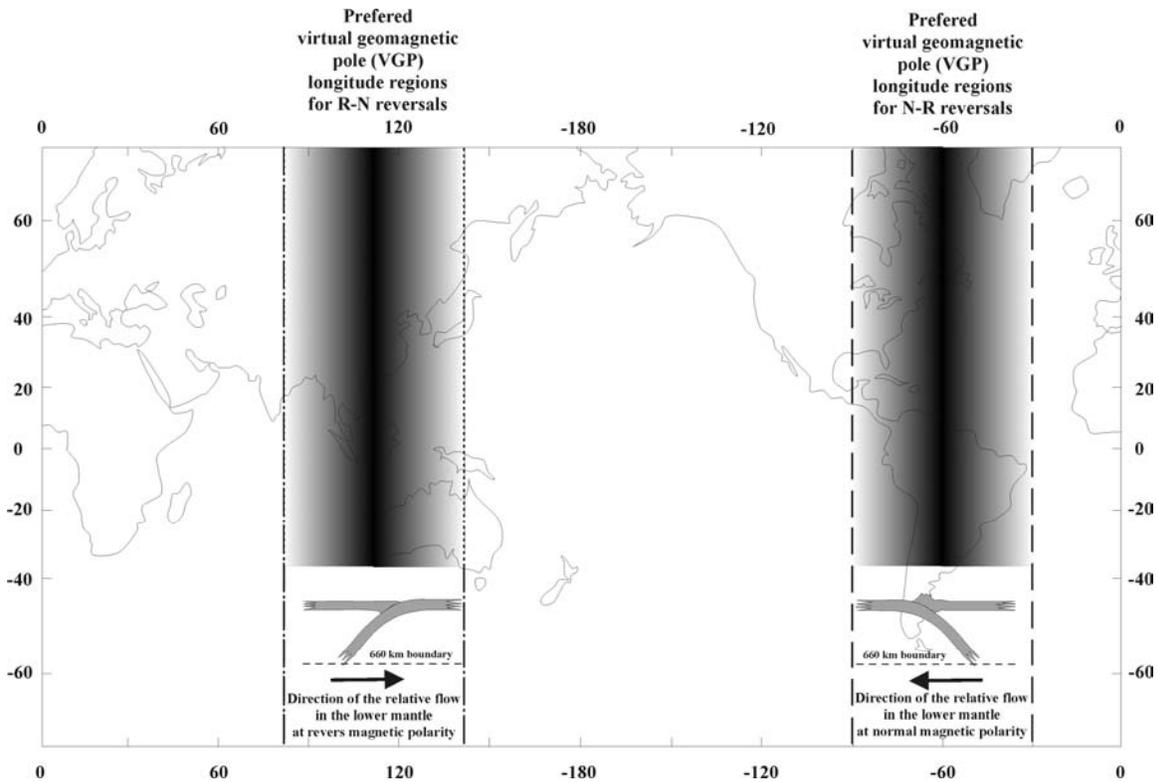

**Figure 5.** The observed preferred longitudes for the virtual geomagnetic poles and the proposed physical explanation for their locality are shown below. If the polarity is reverse then the lower mantle rotates faster while at normal polarity the lower mantle rotes slower. The sign of the rotation changes at the most rigid part of the mantle (subduction zones) where the direction of the slab is opposite to the flow.

The impact triggered global cooling resulting an R-N reversal is fast and should generate a strong field right after the reversal. Observations of the past four million years show an asymmetrical saw-tooth patterns and correlation between the amplitude of the field, following the reversals, and the length of the polarity interval (Valet and Meynadier, 1993). On the other hand the N-R reversals do not requires catastrophic events therefore the transition should be smooth and the intensity should not be necessarily high following the reversals. The amplitude of N-R reversal is generally smaller in comparison to R-N reversals. The strength of the normal field depends on the size of the glaciation.  The correlation between the amplitude and the length of the period could be explained by the longer "recovery" period from a sizable glaciation.

Longer term without reversal should develop a higher differential rotation between the outer 660 km and the rest of the earth.  Stronger magnetic fields detected for superchrons are consistent with this prediction.  This higher differential rotation would result a faster movement of the plates, which would produce more volcanic material.  The detected inverse correlation between the frequency of the magnetic reversals and plume and volcanic activity is consistent with this prediction.

The lack of plate tectonics and intrinsic magnetic field on Venus is consistent with the very small angular momentum of the planet.

## 8. Conclusion

Global coupling at 660 km is consistent with plate tectonics and offers a plausible explanation for the generation of the earth's magnetic field.  The reversal of the field occurs when the sign of the differential rotation at 660 km changes.  R-N reversal results from global cooling while N-R reversals occur resulting from tidal drag which might be associated with global worming.

The sign of the rotation shifts first where the resistance between the 660 km outer-shell and the rest of the earth is strongest.  The preferred longitudes for the virtual geomagnetic poles along the subduction zones are strong indicatives that the field is not generated in the inner core.

Assuming spherical distribution of uniform charge, the differential rotation model is able to reproduce all of the features of the contemporary and, within reasonable uncertainty, the paleomagnetic field.  The model is also consistent with geological and geophysical observations correlating to the magnetic field.